\documentclass[sigconf]{acmart}
\copyrightyear{2024}
\acmYear{2024}
\setcopyright{acmlicensed}\acmConference[SIGCSE 2024]{Proceedings of the 55th ACM Technical Symposium on Computer Science Education V. 1}{March 20--23, 2024}{Portland, OR, USA}
\acmBooktitle{Proceedings of the 55th ACM Technical Symposium on Computer Science Education V. 1 (SIGCSE 2024), March 20--23, 2024, Portland, OR, USA}
\acmPrice{15.00}
\acmDOI{10.1145/3626252.3630780}
\acmISBN{979-8-4007-0423-9/24/03}
\usepackage{multirow}
\usepackage{dblfloatfix}
\usepackage{colortbl}
\usepackage{xcolor}
\usepackage{rotating}
\usepackage{enumitem}

\definecolor{giallo}{RGB}{255,153,0}
\definecolor{blu}{RGB}{102,140,217}
\definecolor{jscriptl}{RGB}{153,255,204}

\def\tsc#1{\csdef{#1}{\textsc{\lowercase{#1}}\xspace}}
\tsc{WGM}
\tsc{QE}
\tsc{EP}
\tsc{PMS}
\tsc{BEC}
\tsc{DE}

\begin{document}
\let\WriteBookmarks\relax
\def\floatpagepagefraction{1}
\def\textpagefraction{.001}



\title[Collaborative Learning Research in Software Engineering Education]{Application of Collaborative Learning Paradigms within Software Engineering Education: A Systematic Mapping Study}                      


%


\author{Rita Garcia}
\orcid{0000-0003-4615-4921}
\affiliation{
  \institution{Victoria University of Wellington}
  \city{Wellington} \country{New Zealand}
}
\email{rita.garcia@vuw.ac.nz}

\author{Christoph Treude}
\orcid{0000-0002-6919-2149}
\affiliation{
  \institution{University of Melbourne}
  \city{Melbourne} \country{Australia}
}
\email{christoph.treude@unimelb.edu.au}

\author{Andrew Valentine}
\orcid{0000-0002-8640-4924}
\affiliation{
  \institution{University of Melbourne}
  \city{Melbourne} \country{Australia}
}
\email{andrew.valentine@unimelb.edu.au}




\begin{abstract}
Collaboration is used in Software Engineering (SE) to develop software. Industry seeks SE graduates with collaboration skills to contribute to productive software development. SE educators can use Collaborative Learning (CL) to help students develop collaboration skills. This paper uses a Systematic Mapping Study (SMS) to examine the application of the CL educational theory in SE Education. The SMS identified 14 papers published between 2011 and 2022. We used qualitative analysis to classify the papers into four CL paradigms: Conditions, Effect, Interactions, and Computer-Supported Collaborative Learning (CSCL). We found a high interest in CSCL, with a shift in student interaction research to computer-mediated technologies. We discussed the 14 papers in depth, describing their goals and further analysing the CSCL research. Almost half the papers did not achieve the appropriate level of supporting evidence; however, calibrating the instruments presented could strengthen findings and support multiple CL paradigms, especially opportunities to learn at the social and community levels, where research was lacking. Though our results demonstrate limited CL educational theory applied in SE Education, we discuss future work to layer the theory on existing study designs for more effective teaching strategies. 

\end{abstract}





\maketitle

\section{Introduction} \label{section_introduction}

Collaboration in Software Engineering (SE) often involves conflict resolution, decision-making, problem-solving, and communication skills \cite{webb:1995}, enabling teams to create software applications and services effectively. These skills are desirable to employers \cite{koncz:2022}, but unfortunately, prior research \cite{begel:2008} has shown SE graduates need help with communication and collaboration early in their professional careers. Employers find graduates need more training and believe institutions could do more to help them \cite{gary:2009}. To address the industry's expectations, students can develop and practice collaboration skills through \textit{Collaborative Learning} (CL), an educational theory and pedagogy that educators use to bring students together to master learning concepts \cite{cohen:1994}. CL can guide the instructional design and support SE educators in situating an activity's context \cite{navarro:2009}.


This paper examines the application of the CL educational theory in SE
Education using a Systematic Mapping Study (SMS), a literature review process designed for SE \cite{petersen:2008}. Previous literature reviews reported on Computer-Supported Collaborative Learning (CSCL) \cite{jeong:2019, zheng:2019}, the collaborative learning conditions for large cohorts \citep{manathunga:2015} and student group formation \citep{maqtary:2019}. Our review is across all CL paradigms: Conditions, Effect, Interactions, and CSCL \cite{lai:2011}. We use the SMS to identify papers published between 2011 and 2022 and apply qualitative analysis to classify the results within the CL paradigms to answer the following research questions:


\begin{itemize}[leftmargin=*]
    \item \textbf{RQ1:} \textit{How is the CL educational theory applied in SE Education?}
    \item \textbf{RQ2:} \textit{What are the overall goals for the CL research?}
    \item \textbf{RQ3:} \textit{How does the level of evidence support the papers' goals?}
    \item \textbf{RQ4:} \textit{How is CSCL research applied in SE Education?}
\end{itemize}

The SMS identified 14 papers, confirming a high level of interest in CSCL. The papers focused on improving student learning, supporting educators in administering and monitoring activities, and enhancing curriculum designs. Our findings demonstrated limited research layering the CL educational theory in SE Education. We discuss how researchers might review existing study designs to align the collaboration aspects of their work with the key elements of the theory designed for higher learning. When examining the papers in depth, we observed almost half did not achieve the appropriate level of supporting evidence. However, calibrating the instruments used in these papers could strengthen the findings and address CL paradigms lacking research, such as CSCL’s aim at understanding learning at the social and community layers. 





\begin{figure*}[t!]
  \centering
  \includegraphics[width=.9\linewidth]{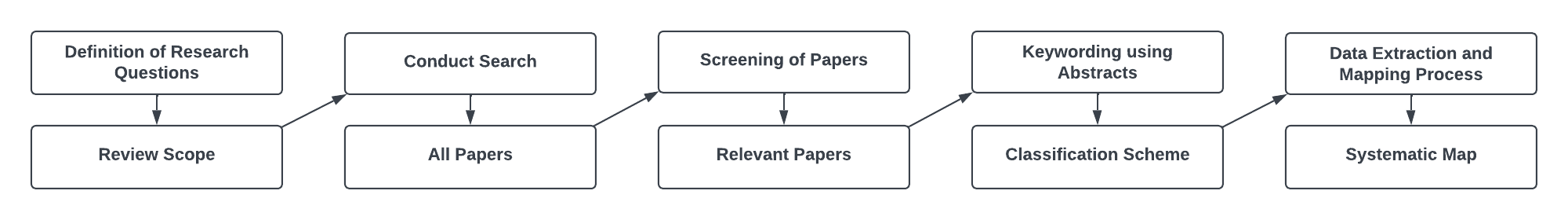}
  \vspace*{-6mm}\caption{Diagram of the Systematic Mapping Study \cite{petersen:2008}}
  \label{fig:data_pipeline_figure}
\end{figure*}

\section{Background} \label{section_background}


\subsection{Students' CL Experiences} \label{section_student_collab_learning_bkgd}


Collaborative Learning (CL) is an educational theory that promotes student engagement through activities that have them working together to form a solution \citep{cohen:1994}. During CL, students share ``goals, symmetry of structure, and a high degree of negotiation, interactivity, and interdependence'' \cite[p.~2]{lai:2011}. CL can be unstructured group work and informal student-teacher interactions, where the teacher employs questioning to encourage dialogue and collaboration \cite{mason:2020}. Collaborative interactions can promote cognitive development \cite{sung:2013} that engages the learning process \cite{dillenbourg:1996} while developing collaboration skills necessary for industry \citep{oneil:1992}.


CL is challenging for students \citep{begel:2008} due to the social pressures that influence group conflicts which impede the learning process \citep{johnson:1979}. Though students find CL challenging, they appreciate collaboration because of the support system it provides for them to ask questions, which helps them to feel less isolated \cite{chang:2018}. Prior collaboration experiences can influence students' expectations, stress levels, and behaviours towards their peers \cite{falkner:2013}. \citet{falkner:2013} observed task-focused students having higher stress levels during a CL activity than those aware of the activity's metacognitive learning objectives, allowing them to concentrate on the team's metacognitive development rather than their performance. 

Educators can assess an individual's contribution to the group activity \citep{hayes:2003} to reduce students' negative experiences during collaboration. Non-contributing group members can create an imbalance in the group's workload \citep{lam:2018}. Fair group assessment might mitigate the effects of non-contributing group members \citep{fincher:2001} to better reflect an individual's accomplishments in the group work \citep{yadav:2020}. \citet{schulz:2022} observed that educators can support students' collaboration skills by providing guidance on collaboration tools. 

The literature presented in this section shows students encountering challenges when collaborating. Collaborative activities also challenge SE students, because they require technical and communication skills that engage individual and social competencies \cite{sedelmaier:2012}. However, the literature also describes ways of mitigating these challenges and improving students' CL experiences.



\subsection{CL Paradigms and Learning Models} \label{section_collab_learning_paradigms}
Collaborative Learning (CL) has been previously classified into four paradigms to understand the depth and focus of the pedagogical approaches \cite{lai:2011}. In this section, we describe these four paradigms\textemdash \textit{Conditions}, \textit{Effect}, \textit{Interactions}, and \textit{Computer-Supported Collaborative Learning (CSCL)}\textemdash along with learning models that are guided by the CL educational theory. 

The first paradigm, \textit{Conditions}, focuses on the environment to promote productive learning. It examines the surroundings and circumstances that influence learning, such as group size \cite{dillenbourg:1996}. The \textit{Effect} paradigm explores how collaboration and the activity's design influence learning. This paradigm considers how student group formation affects learning from students with different abilities. The \textit{Interactions} paradigm examines the interplay between CL processes and the student's abilities that influence learning outcomes. This paradigm relates to the \textit{Conditions} paradigm by evaluating the factors that impact students during the CL process.
    

The last paradigm, \textit{Computer-Supported Collaborative Learning (CSCL)}, investigates how computers can support students' learning. The first three paradigms help ``disentangle the different research paradigms and theoretical approaches'' \cite[p.~8]{dillenbourg:1996}. CSCL emerged later \cite{stahl:2006}, using computers as a medium for communication and collaboration, to support the development of social skills during learning \cite{jeong:2016}. \citet{kirschner:2013} previously developed a 3x3 theoretical framework to conceptualise CSCL research. We use this framework in our study design (See Section \ref{section_rq_method}) and present its dimensions along with definitions in Table \ref{table_cscl_dimensions_results}.
    



Some learning models support educators in practising CL with their students, such as Jigsaw, Problem-Based Learning, and informal and formal CL groups \cite{goodsell:1992}. The Jigsaw model encourages individuals to concentrate on a subset of concepts so that they can explain them to peers \cite{aronson:1997}; this helps the student to become the expert in that area and promotes their self-confidence. Problem-Based Learning (PBL) makes the problem the focus of student discussions, enabling them to form objectives for self-directed learning \cite{wood:2003}. Informal and formal learning groups, such as discussions, promote dialogues between educators and peers to exchange ideas and opinions that improve students' understanding of learning concepts, so they can internalise concepts before proceeding to new ones \cite{felder:2009}. 


In addition to the learning models, instructional approaches also promote CL. For example, Process Oriented Guided Inquiry Learning (POGIL) has students working together while the educator serves as a facilitator \citep{moog:2008}. While the collaboration process might not be the focus of POGIL activities, it encourages students to explore completing tasks together \citep{macgregor:1992}. The literature presented in this section is not an exhaustive list evaluating theoretical paradigms, learning models, and instructional materials. Still, it demonstrates how the CL educational theory has influenced learning environments. We examined practical collaborative approaches to give us a reference for our research on how prior research applies the CL educational theory in SE Education.

\vspace*{-3mm}\section{Methodology} \label{section_rq_method}

We used a Systematic Mapping Study (SMS) to identify papers researching CL in SE Education. SMS is a literature review method for the SE discipline \cite{petersen:2008} that combines guidelines for identifying research trends \cite{petersen:2015}. Prior research \cite{ali:2019} has used SMS to determine interests in mixed reality technology in education. Figure \ref{fig:data_pipeline_figure} shows the ten SMS states. In this section, we discuss four key SMS states that explain how we conducted this study. The files mentioned are available in our online appendix hosted on Figshare.\footnote{https://doi.org/10.6084/m9.figshare.22002542}



\subsection{Review Scope} \label{subsection_scope}

In the \textit{Review Scope} state (See Figure \ref{fig:data_pipeline_figure}), we identify papers of interest within SE courses, workshops, or programs that recruit SE educators or students as participants. We sought peer-reviewed papers published between 2011 and 2022 since 2011 marks the advent of GitHub and Stack Overflow influencing the landscape of software collaboration \cite{asparouhova:2016}. GitHub reached two million repositories in 2011, while Stack Overflow emerged as a popular platform for software developers to ask and answer questions online. 

\subsection{Conduct Search and Screening Processes} \label{section_search_screening_processes}

We apply a search query in the SMS \textit{Conduct Search} state to find relevant papers. We identified papers using the ACM and IEEE Xplore Digital libraries, Springer, Elsevier, ScienceDirect, Taylor \& Francis Online, and SAGE Publication websites. The query used the term \textit{``Collaborative Learning''} in the papers' titles to identify the highest number of relevant papers; we also searched for the terms \textit{``Software Engineering''} and \textit{``Education''} anywhere in the paper. 


The meta-search engine found 44 papers, which we processed in the \textit{Screening of Papers} and \textit{Keywording using Abstracts} SMS states. The primary author conducted the filter process in an Excel spreadsheet which a co-author verified. We excluded fourteen (31.82\%) papers because the research was conducted outside of SE courses or workshops or did not have SE educators or students as participants. For example, a study by \citet{wendel:2013} designed a 3D multiplayer game for students to encourage collaborative behaviour, but the study recruited game-playing students outside of an SE course or workshop. We excluded six (13.64\%) papers because they did not directly relate to the CL paradigm. For example, \citet{wang:2022} evaluated a novel CL approach for intelligent transportation systems unrelated to SE education. We also excluded six (13.64\%) other papers because the terms \textit{``Software Engineering''} or \textit{``Collaborative Learning''} appeared exclusively in the related work or background sections, and we excluded four papers (9.10\%) because they were literature reviews. In total, 30 papers did not meet the selection criteria, resulting in 14 (31.82\%) remaining for analysis. 

\subsection{Data Extraction and Mapping Process} \label{section_data_extraction_mapping_process}

In the \textit{Data Extraction and Mapping Process} state, we compared the 14 papers' attributes, such as study group size and country of origin. We also performed qualitative analysis in this state, to classify the papers in the CL paradigms and CSCL dimensions (Section \ref{subsub_paradigm_coding}) and identify the papers' overall goals (Section \ref{subsub_trends_coding}).

\subsubsection{Classification of CL Paradigms and CSCL Dimensions} \label{subsub_paradigm_coding}
We used thematic content analysis \cite{marshall:1999} to classify the 14 papers within the CL paradigms and the CSCL dimensions, which we previously described in Section \ref{section_background}. Table \ref{table_cscl_dimensions_results} presents the 3x3 CSCL dimensions framework, with definitions. The initial coding framework consisted of the four CL paradigms, the nine CSCL dimensions, and \textit{CSCL - Unknown}, a classification for papers focusing on CSCL that were missing the information required to discern a dimension.  

The primary author classified all the papers within the initial coding framework. Upon completion, the authors discussed the coding process for reliability, including how to classify the papers' use of online tools and platforms, such as forums. The authors agreed that the intention to use technologies was a factor for \textit{CSCL} classification. The authors also discussed the difference between coding papers in the \textit{Effect} paradigm and the CSCL dimensions. The authors decided that if a paper described collaboration within computer-mediated technologies, it would be classified within the \textit{CSCL} paradigm; otherwise it belonged in the CL \textit{Effect} paradigm.

The authors also discussed coding the papers' level of supporting evidence, which generated three additional categories to the coding framework. The first, \textit{Demonstrate}, includes papers with strong evidence collected from the instruments to support the findings. The second, \textit{Claim}, is for papers that focus on a paradigm, but without instruments calibrated for that paradigm. The last, \textit{Mention}, is for papers that mention a paradigm without instruments collecting data for that paradigm.







After the authors' discussion for coding reliability, two co-authors recoded the papers using Excel. The co-authors' recoding results were compared for interrater reliability \cite{cohen:1968} using Cohen's Kappa (k) to measure agreement. The recoding of the papers achieved a kappa of 0.85, considered almost perfect agreement \cite{landis:1977}.



\subsubsection{Classification of Papers' Overall Goals} \label{subsub_trends_coding}

We also classified the papers, evaluating their overall research goals. We used thematic content analysis to form emerging themes from the papers' goals for the coding framework. The primary author analysed all the papers using an Excel spreadsheet, classifying them into three emerging goals: \textit{Curriculum Design}, \textit{Educator Support}, and \textit{Student Learning}. For coding reliability, a co-author used the emerging theme framework to code the papers' goals independently. Upon completion, we compared the coding results for interrater reliability \cite{cohen:1968}, finding 78.57\% agreement, a substantial agreement \cite{landis:1977}. The authors further discussed the coding results until consensus was met, attributing coding discrepancies to the primary author considering papers having one overall goal rather than two.

\begin{table*}[h]
\caption{Results from Systematic Mapping Study (* Educators as Participants)}\label{paper_results}
\begin{tabular}{cp{2.5cm}llcp{10.5cm}}
\small{\textbf{\#}}&\small{\textbf{Publication}} & \small{\textbf{Year}}  & \small{\textbf{Country}} & \small{\textbf{Group}} & \small{\textbf{Description}}\\
\hline
\small{1} & \small{\citet{clarke:2014}} &\small{2014}& \small{USA} & \small{-} & \small{Describes a CL environment for students that contains tools and instructional materials that support software testing practices.}\\ \hline
\small{2}&\small{\citet{coccoli:2011}}& \small{2011} & \small{Italy} & \small{3-4} & \small{Uses Rational Tools, such as Ration Team Concert (RTC), to support collaboration across large cohorts in multiple universities in Italy.}\\ \hline
\small{3} & \small{\citet{danielewicz:2014}} &\small{2014}& \small{Japan} & \small{7}& \small{Describes an activity with customers and students collaborating to develop projects that support students' project management and organisational skills.}\\ \hline
\small{4}&\small{\citet{elmahai:2012}}&\small{2012}& \small{Sudan} & \small{2-5} & \small{Incorporates a forum and wiki within Moodle that enables Sudanese female and male students to collaborate outside the classroom.}\\ \hline
\small{5}&\small{\citet{florez:2021}}&\small{2021} & \small{Columbia} & \small{4} & \small{Describes how Zoom supports student collaboration in an object-oriented programming course, describing group structure and roles.}\\ \hline
\small{6}&\small{\citet{martinez:2016}}&\small{2016}& \small{Australia}&  \small{3\textsuperscript{*}} & \small{Leverages collaboration tools, such as a tablet and interactive tabletop, to support collocated educators in high-level course design courses using design patterns.}\\ \hline
\small{7}&\small{\citet{maruyama:2018}}&\small{2018} & \small{Japan} & \small{4} & \small{Supports the refinement of students' conceptual models through CL, piloted within an object-oriented development course.}\\ \hline
\small{8} & \small{\citet{molina:2014}} & \small{2014}& \small{Spain} & \small{8\textsuperscript{*}} & \small{Evaluates the Collaborative Interactive Application Notation (CIAN) that uses eye tracking and think-alouds to collect participants' experiences with the notation.}\\ \hline
\small{9}&\small{\citet{colin:2017}} &\small{2017}& \small{USA} & \small{5-6} & \small{Describes a framework that assists student collaboration, resulting in better working together with higher individual achievements.}\\ \hline
\small{10} & \small{\citet{popescu:2014}} & \small{2014} & \small{Romania} & \small{4-5} & \small{Describes eMUSE, a Web 2.0 CL platform containing services for students and educators to perform, monitor, and grade CL activities.}\\ \hline
\small{11}&\small{\citet{song:2011}}&\small{2011} & \small{China} & \small{3-5} & \small{Uses CL to support students in developing SE skills by having them work together to create real-world applications.} \\ \hline
\small{12}&\small{\citet{teiniker:2011}}&\small{2011} & \small{Austria} & \small{5-8}& \small{Describes pedagogical approaches to promote authentic learning within an SE course, explaining a blend of strategies that enhanced student learning.}\\ \hline
\small{13} & \small{\citet{xiao:2013}} & \small{2013}& \small{USA} & \small{5-6} & \small{Presents how CL promotes reflective thinking and idea exchange between students.}\\ \hline
\small{14}&\small{\citet{yelmo:2011}}&\small{2011} & \small{Spain} & \small{3-5} & \small{Describes developing and integrating a software tool and a web platform that encourages student collaboration in an SE lab.}\\
\end{tabular}
\end{table*}

\section{Results and Discussion} \label{section_results}

Our SMS identified 14 papers researching CL educational theory in SE Education. Our findings support a previous critique \cite{navarro:2009} on the limited use of educational theories in SE Education, which reviewed learning theories in SE Education and explained the benefits of structuring studies. By applying educational theories, we can discover how students interconnect and envision new research opportunities to advance SE Education that would elevate the student's learning experience. \citet{navarro:2009} explained that though study designs may not apply educational theories, key design elements in these papers sometimes relate to the theories. By having researchers reflect on their study designs within the CL educational theory, these studies can elevate students' learning.



Table \ref{paper_results} presents the 14 papers alphabetically, displaying the publication years, countries, group sizes, and descriptions. The table shows that the studies were conducted in a variety of countries. Thirteen (93\%) papers provided the participant group size ($\bar{x}$=4.81, median=4.63). The largest (n=8) participant group size was in a study \cite{molina:2014} involving educators evaluating a Collaborative Interactive Application Notation (CIAN) that models CL activities. In this section, we further discuss the results within the context of the four research questions.

\begin{table*}[b!]
\caption{Classification of the Collaborative Learning (CL) Paradigms Showing the Level of Supporting Evidence}\label{table_coding_results}
\begin{tabular}{l|>{\raggedright\arraybackslash}p{4cm}|>{\raggedright\arraybackslash}p{4cm}|>{\raggedright\arraybackslash}p{2.8cm}||l}
&\multicolumn{3}{c|}{\textbf{Level of Supporting Evidence}} & \\ \hline
\textbf{Paradigm} &\textbf{Demonstrate} & \textbf{Claim} & \textbf{Mention} & \textbf{Total} \\ 
\hline
\textbf{CSCL} & 5 (25\%) \hspace*{4mm}\cite{elmahai:2012, martinez:2016, maruyama:2018, molina:2014, xiao:2013} & 5 (25\%) \hspace*{4mm}\cite{clarke:2014, coccoli:2011, florez:2021, popescu:2014, yelmo:2011} & 0 & 10 (50\%) \\ \hline

\textbf{Effect} & 4 (20\%) \hspace*{4mm}\cite{clarke:2014, colin:2017, danielewicz:2014, song:2011} & 1 (5\%) \hspace*{5.5mm}\cite{yelmo:2011} & 0 & 5 (25\%) \\ \hline

\textbf{Conditions} & 1 (5\%) \hspace*{5.5mm}\cite{clarke:2014} & 1 (5\%) \hspace*{5.5mm}\cite{teiniker:2011} & 2 (10\%) \hspace*{3mm}\cite{colin:2017, elmahai:2012} & 4 (20\%) \\ \hline

\textbf{Interactions} & 1 (5\%) \hspace*{5.5mm}\cite{colin:2017}& 0 & 0 & 1 (5\%) \\ \hline \hline

\textbf{\textit{Total}} & 11 (55\%) & 7 (35\%) & 2 (10\%) & 20 \\ 
\end{tabular}
\end{table*}




\begin{table*}[b]
\caption{Classification of CSCL Papers using the 3x3 CSCL Theoretical Framework \cite{kirschner:2013}}\label{table_cscl_dimensions_results}
\begin{tabular}{l|c|>{\raggedright\arraybackslash}p{1.4cm}|p{11.4cm}}
\textbf{CSCL Dimension} &\textbf{Total} & \textbf{Papers} & \textbf{CSCL Dimension Definition}\\
\hline
\multicolumn{4}{c}{\cellcolor[gray]{0.8}\textbf{Pedagogical Measures (n=4, 33\%)}} \\ \hline
\textbf{Interactive} & 3 (25\%) &\cite{coccoli:2011, elmahai:2012, martinez:2016} & ``facilitation of the communicative and interactive processes between the collaborating students to support discussion, information sharing, and deliberation among them'' \cite{kirschner:2013} \\ \hline
\textbf{Guiding} & 1 (8\%) & \cite{popescu:2014} & ``guide and direct collaborating students through the learning and collaboration'' \cite{kirschner:2013}\\ \hline
\textbf{Representational} & 0 &&``structure or help students to structure and organize information or knowledge at the cognitive, social, or motivational level'' \cite{kirschner:2013}\\ \hline

\multicolumn{4}{c}{\cellcolor[gray]{0.8}\textbf{Level of Learning (n=3, 25\%)}} \\ \hline
\textbf{Cognitive} & 2 (17\%) & \cite{molina:2014, xiao:2013} & ``knowledge of oneself and in a CL situation about others as cognitive processors'' \cite{kirschner:2013}\\ \hline
\textbf{Motivational} & 1 (8\%) & \cite{popescu:2014} &  ``attitudes, values, predispositions, opinions, beliefs toward what is to be learned, and others within a group or team'' \cite{kirschner:2013} \\\hline
\textbf{Social} & 0 & &Focuses on group-oriented collaboration, empowering the group students to improve self- and group efficacy \cite{kirschner:2013} \\ \hline

\multicolumn{4}{c}{\cellcolor[gray]{0.8}\textbf{Unit of Learning (n=3, 25\%)}} \\ \hline
\textbf{Group/Team} & 2 (17\%) & \cite{florez:2021, xiao:2013} & ``through argumentation and discussion with others, the learner will reach a deeper level of learning'' \cite{kirschner:2013}\\ \hline
\textbf{Individual} & 1 (8\%) & \cite{maruyama:2018} & ``aimed at individual learning gains of collaborating participants'' \cite{kirschner:2013}\\ \hline
\textbf{Community} & 0 && ``large entities like group or communities will learn from CL situations'' \cite{kirschner:2013}\\ \hline

\multicolumn{4}{c}{\cellcolor[gray]{0.8}\textbf{CSCL - Unknown (n=2, 17\%)}} \\ \hline
\textbf{CSCL - Unknown} & 2 (17\%) &\cite{clarke:2014, yelmo:2011}
\end{tabular}
\end{table*}

\subsection{RQ1: How is the CL educational theory applied in SE Education?} \label{results_rq1_section}

Table \ref{table_coding_results} presents the classification of the 14 papers within the CL paradigms. The table organises the papers' classifications within the CL paradigms in descending order, along with the levels of supporting evidence, \textit{Demonstrate}, \textit{Claim}, and \textit{Mention}, to answer RQ3 (See Section \ref{subsection_rq3a_results}). The table shows the ways the papers were classified in the coding framework, with the percentages in the table representing the classification distribution. Four \cite{clarke:2014, elmahai:2012, colin:2017, yelmo:2011} of the 14 papers were classified into multiple categories. For example, the work by \citet{yelmo:2011} describes a pilot program with a web platform and SE tool to support CL in a lab environment. This study examines how technologies influence student learning to enhance the learning experience in the lab further. The SE tool, APIS, enabled students to perform systems modelling on lab assessments. The web platform facilitated student collaboration and allowed lab supervisors to track the progress of projects. As a result, this paper classified in the \textit{CSCL} paradigm for its evaluation of computer-mediated technologies and the \textit{Effect} paradigm for measuring learning outcomes within the lab.


Table \ref{table_coding_results} shows the majority (n=10) of the 14 papers focused on CSCL, which we discuss in-depth by answering RQ4 (See Section \ref{subsection_rq3_results}). The \textit{Effect} paradigm was the next commonly researched paradigm with five papers, followed closely by \textit{Conditions} with four papers. An example of a paper that addresses \textit{CSCL}, \textit{Effect}, and \textit{Conditions} is by \citet{clarke:2014}, which evaluates an online repository, \textit{Web-Based Repository of Software Testing Tools} (WReSTT V2) over a three-year period. WReSTT V2 contains tutorials for testing skills and techniques using the CppUnit and JUnit frameworks that students can apply in group projects. \citet{clarke:2014} evaluates WReSTT V2 in under- and graduate courses to measure improvement in students' testing knowledge. 


\textit{Interactions} was the least researched paradigm. The one paper \cite{colin:2017} examines the Online Mediated Collaboration Model (OMCM), a framework using team-building activities for students to get acquainted. The study showed that the framework improves student collaboration but does not influence individual achievements. A potential reason for the limited focus on \textit{Interactions} might be the shift in focusing on student interactions within computer-mediated technologies, which we discuss by answering RQ4 in Section \ref{subsection_rq3_results}.

\subsection{RQ2: What are the overall goals for the CL research?} \label{results_rq2_section}



We identified three categories for the papers' overall research goals: \textit{Student Learning} (9 papers, 56\%), \textit{Educator Support} (4 papers, 25\%), and \textit{Curriculum Design} (3 papers, 19\%). The results show two \cite{clarke:2014, popescu:2014} of the 14 papers having two goals that focus on \textit{Educator Support} and \textit{Student Learning} using CSCL tools. For example, the previously mentioned study by \citet{clarke:2014} on the WReSTT V2 repository supports students' testing skills while aiding educators in creating student teams. \citet{popescu:2014} presents eMUSE, a platform that integrates Twitter to promote a sense of community with the students and enables educators to monitor students. Both studies apply computer-mediated technologies that support students and educators.


The results show nine papers with \textit{Student Learning} as an overall goal, strengthening prior findings \cite{zheng:2019} that CSCL research is predominately focused on students' learning. Our results include an example of a \textit{Student Learning} paper, the previously mentioned pilot program by \citet{yelmo:2011}, which uses online tools and a web platform to support student collaboration within a lab environment.


Fewer (n=4) papers' goals focused on \textit{Educator Support}, emphasising administrating and developing CL activities. For example, \citet{martinez:2016} evaluate the COllocated COllaborative Design Surface (CoCoDeS) platform that uses design patterns to develop educational materials. Design patterns are descriptions of designs that address software behaviours, objects, visualisations, and interfaces to create higher-quality applications and services \cite{gamma:1994}. This research supports the collocation of educational design so that educators can quickly create high-level courses, including forming student groups and providing guidance on when to distribute activities for higher student learning.

\vspace{-4mm}\subsection{RQ3: How does the level of evidence support the papers' goals?}\label{subsection_rq3a_results}


Table \ref{table_coding_results} organises our findings within three levels of supporting evidence\textemdash\textit{Demonstrate}, \textit{Claim}, and \textit{Mention}\textemdash and displays these findings within their respective four CL paradigms. The table shows the papers next to their level of support. 

The majority (n=11) of the 14 papers demonstrate appropriate supporting evidence, such as \citet{danielewicz:2014}, which investigates how students working with customers can promote higher-quality projects. This study examined adjustments to the course for higher student learning by evaluating self-reflection reports and student-customer emails. The self-reflection reports showed that the students appreciated the interpersonal skills introduced in the course.

Some (n=7) of the papers did not demonstrate the appropriate level of evidence for their goals. For example, the previously mentioned work with eMUSE supports the ``learning needs of digital native students'' \cite[p.~201]{popescu:2014}. The paper presents five post-questionnaire questions mainly focused on student's perceptions of the platform, such as \textit{``Were you satisfied with the eMUSE platform?''}. Because the data collected does not focus on the targeted construct, these results could be considered a limitation to the study's content validity, where the assessment instrument has limited domain relevance \cite{rusticus:2014}. Calibrating the instrumentation by demonstrating learning gains through grades or self-reflection reports could bring the data closer to supporting the goal of eMUSE to help students' learning needs.


While discussing their findings, two \cite{colin:2017, elmahai:2012} papers mentioned the CL \textit{Conditions} paradigm. For example, the study by \citet{colin:2017} designs and evaluates a framework that promotes student collaboration. Though this study focuses on the \textit{Effect} and \textit{Interactions} paradigms, they mention the \textit{Conditions} for learning activities, suggesting ``activities should be included only where they are most meaningful and they must be designed and facilitated carefully to ensure that students
not only find value in the team experience'' \cite[p.~600]{colin:2017}. Perhaps the authors reflected on teaching strategies, evaluating how they can advance the research through instrument adjustments. The authors' considerations on activity facilitation to improve learning outcomes is a goal of educational learning theories that ``can be used to categorize, design, evaluate, and communicate about software engineering educational approaches, providing a structured and informed way to move our domain forward with approaches that are effective and well-understood'' \cite[p.~18]{navarro:2009}.


\subsection{RQ4: How is the CSCL research applied in SE Education?}\label{subsection_rq3_results}

In response to RQ1, we found that CSCL was the most commonly researched paradigm, with ten papers in this category. A potential reason for these findings might relate to technological advancements \cite{zheng:2019}, where researchers are increasingly using technology-supported CL, generating a ``perpetual search for the next new technology that can improve and even transform teaching and learning'' \cite[p.~15]{jeong:2019}. The increased interest in CSCL has also promoted literature reviews focusing on this paradigm in STEM \cite{jeong:2019} and learning environments \cite{chen:2018}.

Table \ref{table_cscl_dimensions_results} shows the coding results for the ten papers within the CSCL dimensions. The table shows twelve classifications for the ten papers. The results show two \cite{popescu:2014, xiao:2013} papers focusing on multiple CSCL dimensions, such as \citet{xiao:2013} placed in the \textit{Group/Team} and \textit{Cognitive} dimensions because of its use of student interviews and a survey to measure group and cognitive learning. The table also shows two \cite{clarke:2014, yelmo:2011} papers that could not be classified, since more information was needed for classification in a CSCL dimension.

The table shows the \textit{Representational}, \textit{Social}, and \textit{Community} learning dimensions without research. We cannot speculate on these findings but observe papers we reviewed could address them. For example, \citet{popescu:2014} describes the learning platform eMUSE as supporting social interactions through Twitter and providing a sense of community for the students; however, the study focuses on the individual's learning. eMUSE creates circumstances to research learning through social and community situations, but a missed opportunity with the existing study design. 


When evaluating the CSCL dimensions, we observed \textit{Interactive} having the most research interest with three papers. These results may explain why the limited focus on CL \textit{Interactions} presented in Section \ref{results_rq1_section} due to the shift in using computer-mediated technologies for student interactions.

\section{Limitations} \label{subsection_limitation}

There are limitations to this study. Firstly, we designed our selection criteria to prevent selection bias that would influence our findings and avoid a sample selection that did not accurately reflect CL research within SE Education. Unfortunately, ``the range of practices included within collaborative learning is very wide and the adoption of that terminology does not aid clarity'' \cite[p.~414]{boud:1999}, which could contribute to papers absent from the search results. Additionally, SE Education has ``a major challenge to integrate applied methodology and theory into the practice of software development instead of learning concepts and methodologies as abstract ideas'' \cite[p.~1]{dahiya:2010}. This challenge could result in researchers describing CL with an array of terms, potentially motivating other CL literature reviews \cite{ali:2019, zheng:2019} to use various search terms to widen their scope. While these limitations exist, our study serves as an important, evidence-based starting point for discussing how Collaborative Learning can be effectively applied in Software Engineering Education.

\section{Conclusion and Future Work} \label{section_conclusion_future_work}




This paper focuses on a Systematic Mapping Study (SMS) that presents 14 papers researching the CL educational theory in SE Education. The results show the CSCL paradigm was the most commonly researched, and the focus on student interactions has shifted to using computer-mediated technologies. The results show student learning was the most common goal for these papers, and it confirms previous CL literature reviews. However, our review also shows interest in curriculum design and educator support.

Our results demonstrate the limited application of the CL educational theory within SE Education. Our findings encourage practical collaborative applications informed by theory in this discipline. Future research can examine how existing practical collaboration approaches could layer theory on study designs, creating ``more effective teaching strategies that are rooted in educational theory'' \cite[p.~19]{navarro:2009}. We observed that almost half of the papers did not have the appropriate level of supporting evidence but could be strengthened through instrument calibration. Researchers have future opportunities to widen their scope to address areas we found that had no research, such as the CSCL dimensions promoting learning through the social and community situations.


\bibliographystyle{cas-model2-names}

\bibliography{cas-refs}


\end{document}